\documentclass{appolb}
\usepackage{graphicx}

\begin{document}
\title{Approximate solution of the pairing Hamiltonian in the Berggren basis%
\thanks{Presented at the XXXIV Mazurian Lakes Conference on Physics, Piaski, Poland, September 6-13, 2015}%
}
\author{Alexis Mercenne, Nicolas Michel, Marek P{\l}oszajczak
\address{Grand Acc\'el\'erateur National d'Ions Lourds (GANIL), CEA/DSM - CNRS/IN2P3, BP 55027, F-14076 Caen Cedex, France}
}
\maketitle
\begin{abstract}
 We derive the approximate solution for the pairing Hamiltonian in the Berggren ensemble of single particle states including bound, resonance and non-resonant scattering states. 
\end{abstract}
  
\section{Introduction}
A deeper understanding of the mathematical foundation of phenomenological models of the atomic nucleus is often related to the studies in simple models for which rigorous solutions can be obtained. 
In the domain of weakly bound or unbound exotic nuclei, we are missing such models, even though it is well known that the coupling to continuum may enhance the pairing correlations and the stability of weakly-bound nuclei \cite{michel_2003}. 
 
The objective of this Letter is to derive an approximate solution for the schematic pairing Hamiltonian in the single-particle (s.p.) space involving both bound and continuum states. 
This model is a generalization of the rational Richardson-Gaudin (RG) model \cite{richardson63_1,richardson64_2} in the Berggren ensemble \cite{berggren68_11} of s.p. states. 
We shall generalize the solution found by Richardson for the rational RG model to describe the pairing interaction of particles in bound and continuum s.p. states. 

\section{Pairing Hamiltonian in the Berggren basis}
{
The pairing Hamiltonian for the rational RG model is given by:
\begin{equation}
	H = \sum_{ \alpha }^{ D } { \epsilon }_{ a } { c }_{ \alpha }^{ \dagger } { c }_{ \alpha } + \frac{ G }{ 4 } \sum_{ \alpha , \beta }^{ D } { c }_{ \alpha }^{ \dagger } { c }_{ \bar{\alpha} }^{ \dagger } { c }_{ \bar{\beta} } { c }_{ \beta }
	\label{pairing_hamiltonian}
\end{equation}
where ${\epsilon }_{ a }$ and ${ G }$ are the energies of (bound) s.p. states and the pairing strength, respectively. 
Here, ${ D }$ stands for the number of s.p states, i.e ${ D = 2N + \nu }$ with ${ N }$ the number of pairs, ${ \nu = \sum_{a}^{U} { \nu }_{ a } }$ the number of unpaired particles, and ${ U }$ the number of levels.
The operators ${ { c }_{ \alpha }^{ \dagger }({ c }_{ \alpha }) }$ stand for the particle creation (annihilation) operators and ${ \alpha \equiv \{ a , { m }_{ \alpha } \} = \{ { n }_{ a } , { \ell }_{ a } , { j }_{ a } , { m }_{ \alpha } \} }$. 
	
It is convenient to rewrite this Hamiltonian in terms of pair creation (annihilation) operators ${ { b }_{ a }^{ \dagger } ({ b }_{ a }) }$ and particle number operator ${ { \hat{ n } }_{ a } }$ :
\begin{equation}
	{ \hat{ n } }_{ a } = \sum_{ { m }_{ \alpha } } { c }_{ \alpha }^{ \dagger } { c }_{ \alpha } \;\;\; , \;\;\; { b }_{ a }^{ \dagger } = 1/2 \sum_{ { m }_{ \alpha } } { c }_{ \alpha }^{ \dagger } { c }_{ \bar{\alpha} }^{ \dagger } = { ({ b }_{ a }) }^{ \dagger } 
	\label{}
\end{equation}
Then, the Hamiltonian becomes :
\begin{equation}
	H = \sum_{ a }^{ U } { \epsilon }_{ a } { \hat{ n } }_{ a } + G \sum_{ a , a' }^{ U } { b }_{ a }^{ \dagger } { b }_{ a' }~\ ,
	\label{pairing_hamiltonian_with_pairs_operator}
\end{equation}
Expressing the pairing Hamiltonian (\ref{pairing_hamiltonian}) in the complete one-body Berggren basis \cite{berggren68_11}, including bound $(b)$, resonance $(r)$, and non-resonant $(c)$ states, one obtains:
\begin{eqnarray}
	H &=& \sum_{ i \in b,r } { \epsilon }_{ i } { \hat{ n } }_{ i } + \sum_{ c } \int_{ { L }_{ c }^{ + } } { \epsilon }_{ { k }_{ c } } { \hat{ n } }_{ { k }_{ c } }  d { k }_{ c } \nonumber \\ 
	&+& G \sum_{ i , i' \in b,r } { b }_{ i }^{ \dagger } { b }_{ i' } + G \sum_{ c , c' } \int_{ { L }_{ c }^{ + } } { b }_{ { k }_{ c } }^{ \dagger } { b }_{ { k' }_{ c' } } d { k }_{ c } d { k' }_{ c' } \nonumber \\ 
	&+& G \sum_{ (i \in b,r) , c } \int_{ { L }_{ c }^{ + } } \left( { b }_{ { k }_{ c } }^{ \dagger } { b }_{ i } + { b }_{ i }^{ \dagger } { b }_{ { k }_{ c } } \right) d { k }_{ c } 
		\label{new_pairing_hamiltonian}
	\end{eqnarray}
	Note that the Hamiltonian of Eq.(\ref{new_pairing_hamiltonian}) is not exactly the same as in Eq.(\ref{pairing_hamiltonian}), as the latter is defined in the Hilbert space, whereas the former is defined in the rigged Hilbert space, comprising bound, resonant, and scattering states \cite{delamadrid}. 
	In this equation, the summation over ${ c \; (c') }$ represents the summation over different partial waves $(\ell, j)$ until $({ \ell }_{ \rm max } , { j }_{ \rm max} )$. ${ { k }_{ c } }$ is related to the s.p energy: ${ \epsilon }_{ c } = { \hbar }^{ 2 } { k }_{ c }^{ 2 } / 2m$, with ${ m }$ the particle mass. 
The discrete sums run over bound s.p. states and s.p. resonances inside the complex plane between the ${ k }$-contour ${ { L }^{ + } }$ and the real $k$-axis. More about the complete Berggren s.p. basis and its application can be found in Ref. \cite{michel09_8}. 
The operators ${ { b }_{ i , { k }_{ c } }^{ \dagger } ({ b }_{ i , { k }_{ c } }) }$ and ${ { \hat{ n } }_{ i , { k }_{ c } } }$ satisfy the commutation relations:
\begin{eqnarray}
	 \left[ { \hat{ n } }_{ i } , { b }_{ i' }^{ \dagger } \right] &=& 2 { \delta }_{ ii' } { b }_{ i }^{ \dagger } \nonumber \\
	 \left[ { b }_{ i } , { b }_{ i' }^{ \dagger } \right] &=& 2 { \delta }_{ ii' } \left( \frac{ { \Omega }_{ i } }{ 4 } \pm \frac{ { \hat{n} }_{ i } }{ 2 } \right)  
	\label{commutator_discrete}
\end{eqnarray}
for discrete (bound and resonance states) s.p states, and
\begin{eqnarray}
	\left[ { \hat{ n } }_{ { k }_{ c } } , { b }_{ { k' }_{ c } }^{ \dagger } \right] &=& 2 \delta ({ k }_{ c } - { k' }_{ c }) { b }_{ { k }_{ c } }^{ \dagger } \nonumber \\
	\left[ { b }_{ { k }_{ c } } , { b }_{ { k' }_{ c } }^{ \dagger } \right] &=& \delta ({ k }_{ c } - { k' }_{ c } ) \frac{ { \Omega }_{ { k }_{ c } } }{ 2 } \pm { \delta }_{ { k }_{ c } { k' }_{ c } } { \hat{n} }_{ { k }_{ c } }
	\label{commutator_continuum}
\end{eqnarray}
for the non-resonant scattering s.p states.
Here, ${ { \Omega }_{ i , { k }_{ c } } }$ is the degeneracy defined as ${ { \Omega }_{ i , { k }_{ c } } = 2 { j }_{ i , { k }_{ c } } + 1 }$.
A Kronecker delta involving scattering states had to be introduced in Eq.(\ref{commutator_continuum}) to preserve antisymmetry.
The upper (lower) signs in these equations stand for bosons (fermions). 

To discretize continuum in Eq. (\ref{new_pairing_hamiltonian}), it is convenient to define new operators: ${ { \tilde{b} }_{ q }^{ \dagger } = { b }_{ q }^{ \dagger } \sqrt{ { w }_{ q } } }$ and ${ { \hat{ \tilde{n} } }_{ q } = { w }_{ q } { \hat{ n } }_{ q } }$ , where ${ { w }_{ q } }$ is a weight which equals one for bound and resonance states, and is equal to the Gauss-Legendre quadrature weight for scattering states.
Index ${ q }$ runs over bound, resonance and discretized scattering states which are all normalized to one. New operators satisfy: 
\begin{eqnarray}
	&& \left[ { \hat{ \tilde{n} } }_{ q } , { \tilde{b} }_{ q' }^{ \dagger } \right] = 2 { \delta }_{ qq' } { \tilde{b} }_{ q }^{ \dagger } \\ 
	&& \left[ { \tilde{b} }_{ q } , { \tilde{b} }_{ q' }^{ \dagger } \right] = 2 { \delta }_{ qq' } \left( \frac{ { \Omega }_{ q } }{ 4 } \pm \frac{ { \hat{ \tilde{n} } }_{ q } }{ 2 } \right) 
	\label{commutator_new_operator}
\end{eqnarray}
The pairing Hamiltonian expressed in these operators: 
\begin{equation}
	H = \sum_{ q }^{ U } { \epsilon }_{ q } { \hat{ \tilde{n} } }_{ q } + \sum_{ q , q' }^{ U } { G }_{ qq' } { \tilde{b} }_{ q' }^{ \dagger } { \tilde{b} }_{ q } ~~;~~  { { G }_{ qq' } = \sqrt{ { w }_{ q } } \sqrt{ { w }_{ q' } } } G~ \ .
	\label{new_hamiltonian_discretized}
\end{equation}
contains the state-dependent pairing force and is not integrable \cite{dukelsky01_31}.

	}

\section{The approximate solution of the pairing Hamiltonian}
{
An approximate solution for the pairing Hamiltonian (\ref{new_hamiltonian_discretized}) in the continuum can be found by replacing the Kronecker delta in the commutation relation (\ref{commutator_continuum}) by the Dirac delta :
\begin{equation}
	\left[ { b }_{ { k }_{ c } } , { b }_{ { k' }_{ c } }^{ \dagger } \right] = 2 \delta ({ k }_{ c } - { k' }_{ c }) \left( \frac{ { \Omega }_{ { k }_{ c } } }{ 4 } \pm \frac{ { \hat{n} }_{ { k }_{ c } } }{ 2 } \right)
		\label{modified_commutator}
\end{equation}
With this change, the new pair operators ${ { \tilde{b} }_{ q }^{ \dagger } ({ \tilde{b} }_{ q }) }$ satisfy:
\begin{equation}
	\left[ { \tilde{b} }_{ q } , { \tilde{b} }_{ q' }^{ \dagger } \right] = 2 { \delta }_{ qq' } \left( \frac{ { \Omega }_{ q } }{ 4 } \pm \frac{ { \hat{ \tilde{n} } }_{ q } }{ 2 { w }_{ q } } \right)
		\label{modified_commutator_new_operator}
\end{equation}
and the commutation relation involving ${ { \hat{ \tilde{n} } }_{ { k }_{ c } } }$ remains the same.

Let us now derive the solution of a Schr\"{o}dinger equation for fermions:  
\begin{equation}
H |{ \Psi }_{ n;\rm norm } > = { E }_{ n } |{ \Psi }_{ n;\rm norm }> ~~;~~ { { E }_{ n } = \sum_{ \gamma = 1 }^{ N } { E }_{ { J }_{ \gamma } } + \sum_{ q }^{ U } { \nu }_{ q } { \epsilon }_{ q } }
\end{equation}
where ${ { E }_{ { J }_{ \gamma } } }$ are the pair energies. 
The derivation for bosons is similar and will not be given here. 

For the many-body state, we take the following ansatz : 
\begin{equation}
	|{ \Psi }_{ n;\rm norm }\rangle = \prod_{ \gamma = 1 }^{ N } { B }_{ { J }_{ \gamma } ; \rm norm }^{ \dagger } |{ \nu }\rangle ~~;~~ { B }_{ { J }_{ \gamma } ; \rm norm }^{ \dagger } = { c }_{ { J }_{ \gamma } } G \sum_{ q }^{ U } \frac{ { \tilde{b} }_{ q }^{ \dagger } \sqrt{ { w }_{ q } } }{ 2 { \epsilon }_{ q } - { E }_{ { J }_{ \gamma } } }
\end{equation}
where the normalization constants ${ { c }_{ { J }_{ \gamma } } }$ are given by:
\begin{equation}
	\frac{ 1 }{ { ( { c }_{ { J }_{ \gamma } } G ) }^{ 2 } } = \frac{ 1 }{ { { C }_{ { J }_{ \gamma } } }^{ 2 } } = \sum_{ q }^{ U } \frac{ { w }_{ q } }{ { (2 { \epsilon }_{ q } - { E }_{ { J }_{ \gamma } }) }^{ 2 } }
	\label{normalization_constant}
\end{equation}
Here ${ |{ \nu }\rangle }$ is a state of unpaired particles satisfying:
\begin{equation}
	{ \tilde{b} }_{ q } |{ \nu }\rangle = 0 \;\;\; , \;\;\; { \hat{ \tilde{n} } }_{ q } |{ \nu }\rangle = { \nu }_{ q } |{ \nu }\rangle
	\label{}
\end{equation}
To simplify, we define ${ { B }_{ { J }_{ \gamma } }^{ \dagger } = { B }_{ { J }_{ \gamma } ; \rm norm }^{ \dagger } / { C }_{ { J }_{ \gamma } } }$ and write:
\begin{equation}
	| { \Psi }_{ n ; \rm norm } \rangle = \prod_{ \gamma = 1 }^{ N } { C }_{ { J }_{ \gamma } } { B }_{ { J }_{ \gamma } }^{ \dagger } |{ \nu }\rangle = { C }_{ n } | { \Psi }_{ n } \rangle ~~ ; ~~ { { C }_{ n } = \prod_{ \gamma = 1 }^{ N } { C }_{ { J }_{ \gamma } } }
	\label{}
\end{equation}

Let us begin by evaluating the commutator:
\begin{equation}
	\left[ H , \prod_{ \gamma = 1 }^{ N } { B }_{ { J }_{ \gamma } }^{ \dagger } \right] = \sum_{ \gamma = 1 }^{ N } \left( \left( \prod_{ \eta = 1 }^{ \gamma - 1 } { B }_{ { J }_{ \eta } }^{ \dagger } \right) \left[ H , { B }_{ { J }_{ \gamma } }^{ \dagger } \right] \left( \prod_{ \mu = \gamma + 1 }^{ N } { B }_{ { J }_{ \mu } }^{ \dagger } \right) \right)
	\label{commutator_1}
\end{equation}
For that, we rewrite the Hamiltonian (\ref{new_pairing_hamiltonian}) in a discretized form:
\begin{equation}
	H = \sum_{ q }^{ U } { \epsilon }_{ q } { \hat{ \tilde{n} } }_{ q } + G { B }_{ 0 }^{ \dagger } { B }_{ 0 } ~~;~~ { { B }_{ 0 }^{ \dagger } = \sum_{ q }^{ U } { \tilde{b} }_{ q }^{ \dagger } \sqrt{ { w }_{ q } } }~ \ ,
	\label{discretized_hamiltonian}
\end{equation}
where ${ U }$ is the total number of bound, resonance and discretized continuum states. 
Knowing the commutation relations for ${ { \tilde{b} }_{ q }^{ \dagger }, { \tilde{b} }_{ q } }$, and ${ { \hat{ \tilde{n} } }_{ q } }$, one obtains:
\begin{equation}
	\left[ { \hat{ \tilde{n} } }_{ q } , { B }_{ { J }_{ \gamma } }^{ \dagger } \right] = \frac{ 2 { \tilde{b} }_{ q }^{ \dagger } \sqrt{ { w }_{ q } } }{ 2 { \epsilon }_{ q } - { E }_{ { J }_{ \gamma } } }  \;\;\; , \;\;\; \left[ { B }_{ 0 } , { B }_{ { J }_{ \gamma } }^{ \dagger } \right] = \sum_{ q }^{ U } \frac{ { w }_{ q } { \Omega }_{ q } / 2 - { \hat{ \tilde{n} } }_{ q } }{ 2 { \epsilon }_{ q } - { E }_{ { J }_{ \gamma } } }
	\label{commutation_relation_for_demo}
\end{equation}
and
\begin{equation}
	\left[ { \hat{ \tilde{n} } }_{ q } , \prod_{ \mu = \gamma + 1 }^{ N } { B }_{ { J }_{ \mu } }^{ \dagger } \right] = \sum_{ \mu' = \gamma + 1 }^{ N } \frac{ 2 \sqrt{ { w }_{ q } } { \tilde{b} }_{ q }^{ \dagger } }{ (2 { \epsilon }_{ q } - { E }_{ { J }_{ \mu' } }) } \left( \prod_{ \mu = \gamma + 1 ; \neq \mu' }^{ N } { B }_{ { J }_{ \mu } }^{ \dagger } \right)
\label{last_big_commutation_relation}
\end{equation}
Then, using Eq.(\ref{commutation_relation_for_demo}) we get :
\begin{equation}
	\left[ H , { B }_{ { J }_{ \gamma } }^{ \dagger } \right] = { E }_{ { J }_{ \gamma } } { B }_{ { J }_{ \gamma } }^{ \dagger } + { B }_{ 0 }^{ \dagger } \left( 1 + G \sum_{ q }^{ U } \frac{ { w }_{ q } { \Omega }_{ q } / 2 - { \hat{ \tilde{n} } }_{ q } }{ 2 { \epsilon }_{ q } - { E }_{ { J }_{ \gamma } } } \right) \ .
\label{relation}
\end{equation}
Inserting (\ref{relation}) and (\ref{last_big_commutation_relation}) into Eq. (\ref{commutator_1}), one finds:
\begin{eqnarray}
	\left[ H , \prod_{ \gamma = 1 }^{ N_{\rm pair} } { B }_{ { J }_{ \gamma } }^{ \dagger } \right] &=& \sum_{ \gamma = 1 }^{ N_{\rm pair} } { E }_{ { J }_{ \gamma } } \left( \prod_{ \eta = 1 }^{ N_{\rm pair} } { B }_{ { J }_{ \eta } }^{ \dagger } \right) - G \sum_{ \gamma = 1 }^{ N } \sum_{ q }^{ U } \frac{ { B }_{ 0 }^{ \dagger } }{ 2 { \epsilon }_{ q } - { E }_{ { J }_{ \gamma } } } \left( \prod_{ \eta = 1 ; \neq \gamma }^{ N } { B }_{ { J }_{ \eta } }^{ \dagger } \right) { \hat{ \tilde{n} } }_{ q } \nonumber \\
	&+& \sum_{ \gamma = 1 }^{ N } \left( 1 + \sum_{ q }^{ U } \frac{ G { w }_{ i } { \Omega }_{ q } / 2 }{ 2 { \epsilon }_{ q } - { E }_{ { J }_{ \gamma } } } \right) { B }_{ 0 }^{ \dagger } \left( \prod_{ \eta = 1 ; \neq \gamma  }^{ N } { B }_{ { J }_{ \eta } }^{ \dagger } \right) \nonumber \\
	&-& \sum_{\stackrel{ \gamma = 1 }{\mu = \gamma + 1}  }^{ N } \!\!\! \sum_{ q }^{ U } \frac{ 2 G \sqrt{ { w }_{ q } } { B }_{ 0 }^{ \dagger } { \tilde{b} }_{ q }^{ \dagger } }{ ( 2 { \epsilon }_{ q } - { E }_{ { J }_{ \gamma } } ) ( 2 { \epsilon }_{ q } - { E }_{ { J }_{ \mu } } ) } \left( \prod_{ \eta = 1 ; \neq \mu , \gamma }^{ N } { B }_{ { J }_{ \eta } }^{ \dagger } \right)
	\label{big_commutator_final}
\end{eqnarray}
Now, applying the commutator (\ref{big_commutator_final}) on the vacuum state $|{ \nu }\rangle$, and using the relation:
\begin{equation}
	\sum_{ q }^{ U } \frac{ { \tilde{b} }_{ q }^{ \dagger } \sqrt{ { w }_{ q } } }{ (2 { \epsilon }_{ q } - { E }_{ { J }_{ \gamma } }) (2 { \epsilon }_{ q } - { E }_{ { J }_{ \mu } }) } = \frac{ { B }_{ { J }_{ \gamma } }^{ \dagger } - { B }_{ { J }_{ \mu } }^{ \dagger } }{ { E }_{ { J }_{ \gamma } } - { E }_{ { J }_{ \mu } } }
	\label{commutation_relation_2}
\end{equation}
we obtain:		
\begin{eqnarray}
	 H |{ \Psi }_{ n }\rangle &=& { E }_{ n } |{ \Psi }_{ n }\rangle + \sum_{ \gamma = 1 }^{ N_{\rm pair} } \left( 1 + \sum_{ q }^{ {\cal N} } \frac{ G { w }_{ q } \left( { \Omega }_{ q } / 2 - { \nu }_{ q } \right) }{ 2 { \epsilon }_{ q } - { E }_{ { J }_{ \gamma } } } \right) { B }_{ 0 }^{ \dagger } \left( \prod_{ \eta = 1 ; \neq \gamma }^{ N_{\rm pair} }  { B }_{ { J }_{ \eta } }^{ \dagger } \right) |{ \nu }\rangle \nonumber \\
	&-& \sum_{ \gamma = 1 }^{ N_{\rm pair} } \left( \sum_{ \mu = 1 ; \neq \gamma }^{ N_{\rm pair} } \frac{ 2 G }{ { E }_{ { J }_{ \mu } } - { E }_{ { J }_{ \gamma } } } \right) { B }_{ 0 }^{ \dagger } \left( \prod_{ \eta = 1 ; \neq \gamma }^{ N_{\rm pair} } { B }_{ { J }_{ \mu } }^{ \dagger } \right) |{ \nu }\rangle
	\label{schrodinger_equations_2}
\end{eqnarray}
As ${ H |{ \Psi }_{ n }\rangle = { E }_{ n } |{ \Psi }_{ n } }\rangle$, one obtains the generalized Richardson equation for pair energies of the discretized pairing Hamiltonian in the Berggren basis:
\begin{equation}
	1 - \sum_{ q }^{ U } \frac{ G { w }_{ q } \left( { \nu }_{ q } - { \Omega }_{ q } / 2 \right) }{ 2 { \epsilon }_{ q } - { E }_{ { J }_{ \gamma } } } + \sum_{ \mu = 1 ; \neq \gamma }^{ N } \frac{ 2 G }{ { E }_{ { J }_{ \gamma } } - { E }_{ { J }_{ \mu } } } = 0
	\label{richardson_equations}
\end{equation}
In the continuum limit, when the number of discretized scattering states becomes infinite and their weights go to zero, these equations take the form:
\begin{equation}
	1 - \sum_{ i \in b,r } \frac{ 2G { d }_{ i } }{ 2 { \epsilon }_{ i } - { E }_{ { J }_{ \gamma } } } - \sum_{c}^{ { \ell }_{ \rm max} , { j }_{ \rm max} } \int_{ { L }_{ c }^{ + } } \frac{ 2G { d }_{ { k }_{ c } } }{ { \hbar }^{ 2 } { k }_{ c }^{ 2 } / m - { E }_{ { J }_{ \gamma } } } d { k }_{ c } + \sum_{ \mu = 1 ; \neq \gamma }^{ N } \frac{ 2G }{ { E }_{ { J }_{ \gamma } } - { E }_{ { J }_{ \mu } } } = 0
	\end{equation}
with ${ { d }_{ i } = { \nu }_{ i } / 2 - { \Omega }_{ i } / 4 }$.
}

\section{Discussion and conclusions}
{
It is interesting to discuss limiting cases of the generalized Richardson equation (\ref{richardson_equations}). If we deal with a discrete set of bound s.p. levels, the weights $w_q$ are all equal one and Eq. (\ref{richardson_equations}) reduces to the exact Richardson solution for the rational RG model \cite{richardson63_1,richardson64_2}. 
By the same argument, Eq. (\ref{richardson_equations}) provides an exact solution of the pairing model with a continuum treated in the pole approximation, {\it i.e.} neglecting the non-resonant continuum states. 
Finally, Eq. (\ref{richardson_equations}) is also the exact solution of the pairing model in the non-resonant continuum because in this case one may take the same weights $w_q \equiv w$ for all states $q$ and renormalize the pairing strength ${ G' = G w }$. 
At this level we no longer use the Gauss-Legendre quadrature, as we start from the Richardson equations taken at continuum limit, so that it is possible to use another discretization scheme where all weights are equal for simplicity.
In this particular case, the third sum in Eq. (\ref{richardson_equations}) goes to zero and one obtains:
\begin{equation}
	G \int_{} \frac{ { d }_{ k } }{ 2 { \epsilon }_{ k } - { E }_{ i } } dk = 1 \ .
	\label{final_continuum_richardson_equations}
\end{equation}

In a most general case of the rational RG Hamiltonian in the Berggren ensemble of s.p. states, Eq. (\ref{richardson_equations}) provides a reliable approximation of the exact solution in the limit of weak pairing correlations. 
The quantitative comparison between results of Eq. (\ref{richardson_equations}) and exact results of Gamow Shell Model \cite{GSM1,GSM2} for the pairing Hamiltonian in the Berggren basis will be discussed elsewhere \cite{future_paper}. 
}

\vskip 0.3truecm
We wish to thank J. Dukelsky for useful discussions. This work was supported in part by the COPIN and COPIGAL French-Polish scientific exchange programs.


\begin{thebibliography}{50}
	{
	\bibitem{michel_2003} N. Michel, W. Nazarewicz, M. P{\l}oszajczak, and J. Oko{\l}owicz, Phys. Rev. C \textbf{67} (2003) 054311.
	\bibitem{richardson63_1} R. W. Richardson, Phys. Lett. \textbf{3} (1963) 277. 
	\bibitem{richardson64_2} R. W. Richardson and N. Sherman, Nucl. Phys. \textbf{52} (1964) 221. 
	\bibitem{berggren68_11} T. Berggren, Nucl. Phys. A \textbf{109} (1968) 265. 
	\bibitem{delamadrid} R. de la Madrid, Eur. J. Phys. \textbf{26} (2005) 287. 
	\bibitem{michel09_8} N. Michel, W. Nazarewicz, M. P{\l}oszajczak and T. Vertse, J. Phys. G : Nucl. Part. Phys. \textbf{36} (2009) 013101. 
		\bibitem{dukelsky01_31} J. Dukelsky, C. Esebbag and P. Schuck, Phys. Rev. Lett. \textbf{87} (2001) 066403. 
	\bibitem{delft99_3} J. von Delft and F. Braun, Proc. of the NATO ASI "Quantum Mesoscopic Phenomena and Mesoscopic Devices in Microelectronics", Ankara/Antalya, Turkey, Eds. I.O. Kulik and R. Ellialtioglu, Kluwer Academic Publishers, (1999) 361. 
\bibitem{GSM1} N. Michel, W. Nazarewicz, M. P{\l}oszajczak and K. Bennaceur, Phys. Rev. Lett. \textbf{89} (2002) 042502.
\bibitem{GSM2} R. Id Betan, R. J. Liotta, N. Sandulescu and T. Vertse, Phys. Rev. Lett. \textbf{89} (2002) 042501.
\bibitem{future_paper} A. Mercenne, N. Michel, and M. P{\l}oszajczak, to be published.


		}
	
\end{thebibliography}
\end{document}